\documentclass[12pt]{article}
\usepackage{amsmath}
\usepackage{amssymb}
\newsymbol \blackbox 1004
\newcommand{\eh}{\hfill}\newlength{\sperr}

\newenvironment{proof}{{\settowidth{\sperr}{\bf\rm
Proof}%
\par\addvspace{0.3cm}\noindent\parbox[t]{1.3\sperr}
{\bf\rm P\eh r\eh o\eh o\eh f\eh }%
}}{\nopagebreak\mbox{}
$\blackbox$\par\addvspace{0.3cm}}

\def\p{\partial}
\def\a{\alpha}

\def\s{\sigma}
\def\l{\lambda}

\def\wt{\widetilde}
\def\ov{\overline}
\def\p{\partial}

\newtheorem{Pa}{Paper}[section]

\newtheorem{Rk}[Pa]{{\bf Remark}}

\newtheorem{Ee}[Pa]{{\bf Example}}

\newtheorem{Pn}[Pa]{{\bf Proposition}}
\newcommand{\CC}
{{\mathchoice {\setbox0=\hbox{$\displaystyle\rm
C$}\hbox{\hbox
to0pt{\kern0.4\wd0\vrule height0.9\ht0\hss}\box0}}
{\setbox0=\hbox{$\textstyle\rm C$}\hbox{\hbox
to0pt{\kern0.4\wd0\vrule height0.9\ht0\hss}\box0}}
{\setbox0=\hbox{$\scriptstyle\rm C$}\hbox{\hbox
to0pt{\kern0.4\wd0\vrule height0.9\ht0\hss}\box0}}
{\setbox0=\hbox{$\scriptscriptstyle\rm C$}\hbox{\hbox
to0pt{\kern0.4\wd0\vrule height0.9\ht0\hss}\box0}}}}

\title{On a new integrable nonlinear Schr\"odinger equation with a simple external potential
and its explicit solutions}

\author{Alexander Sakhnovich}

\date{}
\begin{document}
\maketitle

{\bf Address}: Fakult\"at f\"ur Mathematik,
Universit\"at Wien, \\
Nordbergstrasse 15,
A-1090 Wien,
Austria \\

e-mail address: al$_-$sakhnov@yahoo.com\\

\begin{abstract}
A  nonlinear Schr\"odinger equation with  external potential $-(t+b)^{-1}$
is considered and its explicit solutions are constructed.
\end{abstract}


\section{Introduction} \label{intro}
\setcounter{equation}{0}

Linearization and explicit solutions of nonlinear differential
equations are classical and permanently gaining  momentum domains
of research. In particular, Riccati equations are actively used
for this purpose (see, for instance, \cite{Ku}).
Here we use function $\psi_0(t)=\frac{1}{2(t+b)}$ that satisfies
equation $(\psi_0)_t=-2\psi_0^2$ to construct  integrable matrix
nonlinear Schr\"odinger equations 
 with a one parameter family of simple potentials (MNSEep):
\begin{equation} \label{0.2}
2u_t=i(u_{xx}+2u u^* u)+f u  \, ({\mathrm{focusing \, MNSEep}}),
\quad f=f(t)=-(t+b)^{-1},
\end{equation}
\begin{equation} \label{0.2'}
2u_t=i(u_{xx}-2u u^* u)+f u \quad ({\mathrm{defocusing \,
MNSEep}}),
\end{equation}
where $u(x,t)$ is an $m_2 \times m_1$ matrix function, $u_t=\p u/ \p t$, and  $f(t)$
is a scalar function. MNSEep include, in particular, the scalar
case and multi-component case.  Various 
nonlinear Schr\"odinger equations with external potentials are used to 
describe deep water and
plasma waves, light pulses, and energy transport (see, for
instance, recent publications \cite{CSoS, KePS, MiG, SuSu} and
references therein).
The author did not meet  integrable equations
(\ref{0.2}) and (\ref{0.2'}) in the literature before.
B\"acklund-Darboux transformation (see \cite{CC,
C1, GeH, Gu, Mar, MS, Mi, RoS, ZM} and references therein) proved
extremely fruitful to construct explicit solutions of the
integrable equations. In this paper we construct explicit solutions of
MNSEep, using a version of B\"acklund-Darboux
transformation introduced in a series of our papers starting from
\cite{SaA1}. The approach proved already quite successful for many
classical ODEs and integrable wave equations (see, for instance,
more references and results in  \cite{GKS6, KaSa, SaA4, SaA6,
SaA10, SaA11, SZ}). Here we show that its modification is
applicable  to non-integrable equations as well (see Proposition
\ref{PnNSL2}).

\section{Matrix  nonlinear Schr\"odinger equation} \label{NLS}
\setcounter{equation}{0}

First we shall define auxiliary linear systems $w_x=G(x,t,\l)w$
and $w_t=F(x,t,\l)w$ for  MNSEep. Put
\begin{equation} \label{a.1}
G(x,t,\l)=-i \l j + \tau (x,t) B,
\end{equation}
\begin{equation} \label{a.2}
F(x,t,\l)=i\Big(\l^2 j+i \l \tau (x,t) B-
\big(j \tau_x (x,t) B- (\tau (x,t) B)^2 j\big)/2 \Big),
\end{equation}
where
\begin{equation} \label{a.3}
j=\left[
\begin{array}{lr} I_{m_1} & 0 \\ 0 & -I_{m_2}\end{array} \right],
\quad
\tau=\left[
\begin{array}{lr} 0 & -u^* \\ u & 0\end{array} \right], 
\end{equation}
$I_p$ is the $p \times p$ identity matrix, and
\begin{equation} \label{a.4}
\l_x=\psi_0(t) , \quad \l_t=-2\psi_0(t) \l \quad (\psi_0(t)=\frac{1}{2(t+b)}).
\end{equation}
Integrability of the nonlinear Schr\"odinger equations (NSE) have been first shown by Zakharov
and Shabat, and one can easily check that
for the spectral parameter $\l$, independent of $x$ and $t$, and for the subcases $B=I_m \quad (m=m_1+m_2)$
and $B=j$ the focusing and defocusing NSE, respectively, are equivalent to the
zero curvature equation $G_t-F_x+(GF-FG)=0$ (see \cite{TF} for historical remarks).
Dependence of $\l$ on $x$ and $t$ given in (\ref{a.4}) is compatible and generates 
the additional term $-\psi_0 \tau B$
in the zero curvature equation. The next proposition follows.
\begin{Pn}
Equations (\ref{0.2}) and (\ref{0.2'}) are equivalent to the zero
curvature equation  $G_t-F_x+(GF-FG)=0$, where $G$ and $F$ are defined
by (\ref{a.1})-(\ref{a.4}), and $B=I_m$ and $B=j$, respectively. 
\end{Pn}

To construct explicit solutions of the integrable NSE
\begin{equation} \label{1.1}
2u_t=i(u_{xx}\pm 2u u^* u),
\end{equation}
the following procedure was proposed in \cite{SaA1}. We fix an
integer $n>0$, two $n \times n$ parameter matrices $A$ and
$S(0,0)=S(0,0)^*$ and one $n \times m$ ($m=m_1+m_2$) parameter
matrix $\Pi(0,0)$ such that
\begin{equation} \label{1.2}
AS(0,0)-S(0,0)A^*=i\Pi(0,0)B \Pi(0,0)^*, \quad B={\mathrm{diag}}
\{ b_1,  \ldots , b_m \}, \, b_k=\pm 1.
\end{equation}
Here diag means diagonal matrix, $b_k$ are diagonal entries of
$B$.

Next, matrix function $\Pi(x,t)$ is defined by its value
$\Pi(0,0)$ at $x=0$, $t=0$ and equations
\begin{equation} \label{1.3}
\Pi_x=i A \Pi j, \quad \Pi_t=-i A^2 \Pi j.
\end{equation}
 By (\ref{1.3})
matrix function $\Pi$ is a "generalized" eigenfunction of the
auxiliary to NSE (\ref{1.1}) linear systems. Eigenfunctions of
these systems are essential in B\"acklund-Darboux theory (see
\cite{SaA11}, p.710 for a more detailed discussion on
"generalized" eigenfunctions that we use in our approach). Matrix
function $S(x,t)$ is defined by its value $S(0,0)$ and derivatives
\begin{equation} \label{1.3'}
S_x=- \Pi B j  \Pi^*, \quad S_t=A \Pi B j \Pi^*+\Pi B j \Pi^*A^*.
\end{equation}
By (\ref{1.2}) and (\ref{1.3'}) $\Pi_{xt}=\Pi_{tx}$ and
$S_{xt}=S_{tx}$, i.e., compatibility conditions are fulfilled. In
view of (\ref{1.2})-(\ref{1.3'}) the identity
\begin{equation} \label{1.4}
AS(x,t)-S(x,t)A^*=i\Pi(x,t)B \Pi(x,t)^*
\end{equation}
holds, that is matrices $A$, $S$, and $\Pi$ form a so called
$S$-node (see \cite{SaL3} and references in this book on $S$-node
theory and its applications). Indeed, the derivatives of both
parts of (\ref{1.4}) coincide, and so (\ref{1.2}) implies
(\ref{1.4}).
\begin{Pn} \label{PnLAA}  Suppose identity (\ref{1.2}) holds
and matrix functions $\Pi$ and $S$ satisfy equations (\ref{1.3})
and (\ref{1.3'}), respectively. Then, in the points of
invertibility of $S$, the matrix function
\begin{equation} \label{d1}
\tau(x,t):=\big(\xi(x,t)j-j \xi(x,t)\big), \quad
\xi(x,t):=\Pi(x,t)^*S(x,t)^{-1}\Pi(x,t)
\end{equation}
satisfies equation
\begin{equation} \label{1.6}
\tau_{xx}=2(i j \tau_{t}+B(B\tau)^3).
\end{equation}
\end{Pn}
Notice that $\tau$ has the  block
structure given in (\ref{a.3}), where, after partitioning $\Pi$ into two blocks $\Pi=[\Phi_1
\quad \Phi_2]$, we have
\begin{equation} \label{1.5}
u(x,t)=2 \Phi_2(x,t)^*S(x,t)^{-1}\Phi_1(x,t).
\end{equation}
Here $\Phi_k$ consists of $m_k$ columns
($k=1,2$).
Thus formula (\ref{1.6}) implies that $u$ of the form (\ref{1.5}) satisfies focusing NSE
$2u_t=i(u_{xx}+ 2u u^* u)$, when $B=I_m$, and defocusing NSE
$2u_t=i(u_{xx}- 2u u^* u)$, when $B=j$. 

\begin{proof} of Proposition \ref{PnLAA}. A proof of the subcase
$B=I_m$ was given in \cite{SaA1}, Section 2, and the scheme
remains the same, when $B \not=I_m$. According to the first
relations in (\ref{1.3}) and (\ref{1.3'}) we have
\begin{equation} \label{1.8}
\xi_x=\Big(\Pi^*S^{-1}\Pi\Big)_x=i\Big(\Pi^*S^{-1}A\Pi j - j
\Pi^*A^*S^{-1}\Pi\Big)+\Pi^*S^{-1}\Pi B j \Pi^*S^{-1}\Pi.
\end{equation}
By (\ref{1.4}) auxiliary formulas
\begin{equation} \label{d3}
S^{-1}A=A^*S^{-1}+i S^{-1}\Pi B \Pi^*S^{-1}, \quad
A^*S^{-1}=S^{-1}A-i S^{-1}\Pi B \Pi^*S^{-1}
\end{equation}
are immediate. According to (\ref{1.3}), (\ref{1.3'}), and
(\ref{d3}) we get
\begin{equation} \label{d5}
\Big(\Pi^*S^{-1}\Big)_x=-i j \Pi^*A^*S^{-1}+\xi B j \Pi^*S^{-1}=-i
j \Pi^*S^{-1}A+(\xi j - j \xi)B \Pi^*S^{-1}.
\end{equation}
 From (\ref{1.8}) and (\ref{d5}) it follows that
\[
\xi_{xx}=K+K^*, \quad K=j\Pi^*S^{-1}A^2 \Pi j-\Pi^*S^{-1}A^2
\Pi+i(\xi j - j\xi)B\Pi^*S^{-1}A \Pi j
\]
\begin{equation} \label{d4}
+\Big(i\big(\Pi^*S^{-1}A\Pi j - j \Pi^*A^*S^{-1}\Pi\big)+\xi B j
\xi \Big)B j \xi.
\end{equation}
In view of  the second relations in (\ref{1.3}) and (\ref{1.3'})
we have
\begin{equation} \label{d6}
\xi_t=ij \Pi^*\big(A^*\big)^2 S^{-1}\Pi- i \Pi^*S^{-1}A^2 \Pi j
-\Pi^*S^{-1}(A\Pi B j \Pi^*+\Pi B j \Pi^*A^*)S^{-1}\Pi .
\end{equation}
Equation (\ref{1.6}) for $\tau=(\xi j-j \xi )$ follows by simple
calculations from (\ref{d4}) and (\ref{d6}) using auxiliary
equalities (\ref{d3}).
\end{proof}

The non-isospectral case with $\l$ depending rationally
 $x$ and $t$ have been treated in \cite{SaA4}. In particular,
for $\l$ satisfying equations
\[
\l_x(x,t)=\sum_{p=0}^k \psi_p(x,t)\l(x,t)^p, \quad \l_t(x,t)=
\sum_{p=0}^{\wt k} \wt \psi_p(x,t)\l(x,t)^p,
\]
 one requires
\[
A_x(x,t)=\sum_{p=0}^k \psi_p(x,t)A(x,t)^p, \quad A_t(x,t)=
\sum_{p=0}^{\wt k} \wt \psi_p(x,t)A(x,t)^p.
\]
In this paper we use the scheme from \cite{SaA4} to construct
solutions of  MNSEep. For instance, put
\begin{equation} \label{1.7}
A(x,t)=\psi_0(t)(A_0+x I_n), \quad \psi_0(t)=\frac{1}{2(t+b)} \,\, (b=\ov b), 
\end{equation}
i.e.,
\begin{equation} \label{1.7.0}
 A_x=\psi_0 I_n, \, A_t=-2 \psi_0 A.
\end{equation}
The last relation in (\ref{1.7}) follows from the equality
$(\psi_0)_t=-2\psi_0^2$. By (\ref{1.7}) equations (\ref{1.3}) on
$\Pi$ are compatible, i.e.,  condition $\Pi_{xt}=\Pi_{t x}$ is
true.

We modify also  equations (\ref{1.3'}) into
\[
S_x=- \Pi B j \Pi^*, \quad S_t=A \Pi B  j \Pi^*+\Pi B j
\Pi^*A^*-\wt \psi_1 S
\]
\begin{equation} \label{1.7'}
=A \Pi B j \Pi^*+\Pi B j \Pi^*A^*+2 \psi_0 S,
\end{equation}
so that the condition  $S_{xt}=S_{t x}$ holds, and derivatives of
the both sides of (\ref{1.4}) coincide again. Therefore identity
(\ref{1.4}) remains valid.

\begin{Pn} \label{PnNSLep} Suppose identity (\ref{1.2}) holds,
$B=I_m$, and matrix functions $\Pi$, $A$, and $S$ satisfy
equations (\ref{1.3}), (\ref{1.7}), and (\ref{1.7'}),
respectively. Then the matrix function $u(x,t)$ given by
(\ref{1.5}) satisfies MNSEep (\ref{0.2}).
\end{Pn}
\begin{proof}. Denote the expression on the right hand side of
the first equality in (\ref{d4}) by $R_0$ ($R_0=K+K^*$). It
follows from (\ref{1.8}) that
\begin{equation} \label{1.9}
\Big(\Pi^*S^{-1}\Pi\Big)_{xx}=R_0 + i\Big(\Pi^*S^{-1}A_x \Pi j - j
\Pi^*A_x^*S^{-1}\Pi\Big),
\end{equation}
where  the second term on the right hand side of  (\ref{1.9}) is
an additional term that appears, when $A_x \not= 0$. Taking into
account (\ref{1.7}), (\ref{1.9}), and the definition of $\tau$ in
(\ref{d1}) we get
\begin{equation} \label{1.10}
\tau_{xx}=(R_0 j - j R_0)+2 i \psi_0 \tau j
\end{equation}
Denote the expression on the right hand side of (\ref{d6}) by $\wt
R_0$. According to the second relations in (\ref{1.3}) and
(\ref{1.7'}) we have
\begin{equation} \label{1.10'}
\Big(\Pi^*S^{-1}\Pi\Big)_t=\wt R_0-2 \psi_0 \Pi^*S^{-1}\Pi.
\end{equation}
Here $-2 \psi_0 \Pi^*S^{-1}\Pi$ is an additional term that
appears, when $A_t \not= 0$, because the expression for $S_t$ is
modified. Proposition \ref{PnLAA} for the $B=I_m$  case takes the
form
\begin{equation} \label{1.10''}
R_0 j - j R_0=2\big(i j(\wt R_0 j - j \wt R_0)+\tau^3\big).
\end{equation}
Therefore from (\ref{1.10})-(\ref{1.10''})  it follows that
\begin{equation} \label{1.11}
\tau_{xx}=2(i j \tau_{t}+\tau^3)-2 i \psi_0 \tau j.
\end{equation}
From (\ref{a.3}) and (\ref{1.11}) formula (\ref{0.2}) is immediate.
\end{proof}
\begin{Rk} Notice that the block columns $\Phi_k$ of $\Pi=[\Phi_1 \quad
\Phi_2]$, i.e., solutions of (\ref{1.3}), where $A$ satisfies
(\ref{1.7}), are given explicitly by the formula
\begin{equation} \label{1.12}
\Phi_k(x,t)= \big( \exp (-1)^{k+1}\a(x,t) \big)g_k,  \quad
\a(x,t):=i\psi_0(t)\big(\frac{x^2}{2}I_n+x A_0+\frac{1}{2}A_0^2
\big).
\end{equation}
By (\ref{1.3}) and (\ref{1.7}) for $S$ defined via (\ref{1.7'})
the compatibility condition $S_{xt}=S_{tx}$ is satisfied.
Moreover, when $\s(A_0) \cap \s(A_0^*)=\emptyset$ ($\s$ means
spectrum), the matrix function $S$ is uniquely recovered from
(\ref{1.4}). Therefore solution $u$ is constructed in Proposition
\ref{PnNSLep} explicitly.
\end{Rk}
\begin{Ee} Let us consider the simplest example $n=1$,
$A_0=a$ ($a \not= \ov a$), $m_1=m_2=1$ in a more detailed way.
For this case formulas (\ref{1.4}), (\ref{1.5}), (\ref{1.7}) and (\ref{1.12})
yield:
\begin{equation} \label{a.5}
u(x,t)=\frac{(a - \ov a)g_1\ov{ g_2}(t+b)^{-1}\exp\Big(\a(x,t)- \ov{\a(x,t)}\Big)}
{i\Big(|g_1|^2\exp\Big(\a(x,t)+ \ov{\a(x,t)}\Big)+|g_2|^2\exp\Big(-\a(x,t)- \ov{\a(x,t)}\Big)
\Big)},
\end{equation}
where
\begin{equation} \label{a.6}
\a(x,t) - \ov{\a(x,t)}=i \psi_0(t)\Big(x^2+x(a + \ov a)+ \frac{1}{2}(a^2 + \ov a^2)\Big).
\end{equation}
\begin{equation} \label{a.7}
\a(x,t)+ \ov{\a(x,t)}=i \psi_0(t)\Big(x(a - \ov a)+ \frac{1}{2}(a^2 - \ov a^2)\Big),
\end{equation}
\end{Ee}
Similar to Proposition \ref{PnNSLep}, our next proposition can be proved.
\begin{Pn} \label{PndfNSLep} Suppose identity (\ref{1.2}) holds,
$B=j$, and matrix functions $\Pi$, $A$, and $S$ satisfy equations
(\ref{1.3}), (\ref{1.7}), and (\ref{1.7'}), respectively. Then the
matrix function $u(x,t)$ given by (\ref{1.5}) satisfies defocusing
MNSEep (\ref{0.2'}).
\end{Pn}

We can choose more complicated equations on $A$ as, for instance:
\begin{equation} \label{1.13}
 A_x=\psi_1 A, \, \, A_t=-2 \psi_1 A^2, \quad
 \psi_1=\psi_1(x)=(x+b)^{-1}, \, {\mathrm{i.e.}}, \,
 (\psi_1)_x=-\psi_1^2.
\end{equation}
By (\ref{1.13}) the compatibility conditions for
the first two relations in (\ref{1.3}) and also in (\ref{1.13})
are satisfied. Moreover the equations
\begin{equation} \label{1.14}
S_x=- \Pi j \Pi^*-\psi_1 S \quad S_t=A \Pi j \Pi^*+\Pi j
\Pi^*A^*+2 \psi_1 (AS+SA^*)
\end{equation}
on $S$ are compatible as well, and yield identity (\ref{1.4}).

\begin{Pn} \label{PnNSL2} Suppose identity (\ref{1.2}) holds,
$B=I_m$, and matrix functions $\Pi$, $A$, and $S$ satisfy
equations (\ref{1.3}), (\ref{1.13}), and (\ref{1.14}),
respectively. Then the matrix function $\xi=\Pi^*S^{-1}\Pi$
satisfies equation
\[
\psi_1^2(j \xi -\xi j)+\psi_1( \xi_x j -j \xi_x )+( \xi_{xx} j -j
\xi_{xx} )+4 \psi_1(j \xi j \xi-\xi j \xi j)
\]
\begin{equation} \label{1.15}
=2\big(i j ( \xi_t j -j \xi_t )+( \xi j -j \xi  )^3 \big).
\end{equation}
\end{Pn}
\begin{proof}.
The scheme of the proof is similar to the proof of Proposition
\ref{PnNSLep}. Instead of (\ref{1.8}) formulas (\ref{1.3}) and
(\ref{1.14}) imply
\begin{equation} \label{1.16}
\xi_x=i\Big(\Pi^*S^{-1}A\Pi j - j \Pi^*A^*S^{-1}\Pi\Big)+\xi j \xi
+\psi_1 \xi .
\end{equation}
After simple calculations from (\ref{1.3}), (\ref{1.13}),
(\ref{1.14}), and (\ref{1.16}) it follows that
\begin{equation} \label{1.18}
\xi_{xx}=R_0+3\psi_1 \xi_x -3  \psi_1^2 \xi.
\end{equation}
(Compare the formula above with (\ref{1.9}).) By (\ref{1.18}) we
have
\begin{equation} \label{1.19}
R_0 j -j R_0=3 \psi_1^2( \xi j - j \xi )-3 \psi_1( \xi_x j -j
\xi_x )+( \xi_{xx} j -j \xi_{xx} ).
\end{equation}
Using (\ref{1.3}) and (\ref{1.14}) we also get
\[
\xi_{t}=\wt R_0-2\psi_1 \big(\Pi^*S^{-1}A\Pi + \Pi^*A^*S^{-1}\Pi
\big), \quad \wt R_0j-j\wt R_0=\xi_{t}j - j \xi_{t}
\]
\begin{equation} \label{1.20}
 + 2\psi_1\Big(\big(\Pi^*S^{-1}A\Pi +
\Pi^*A^*S^{-1}\Pi \big)j-j\big(\Pi^*S^{-1}A\Pi + \Pi^*A^*S^{-1}\Pi
\big)\Big).
\end{equation}
According to  (\ref{1.16}) we rewrite the terms with $A$ on the
right hand side of the second equality in (\ref{1.20}):
\[
\big(\Pi^*S^{-1}A\Pi + \Pi^*A^*S^{-1}\Pi
\big)j-j\big(\Pi^*S^{-1}A\Pi + \Pi^*A^*S^{-1}\Pi \big)
\]
\begin{equation} \label{1.21}
=-i(\xi_x j - j \xi_x)j+i \psi_1( \xi j - j \xi )j+i(\xi j \xi-j
\xi j \xi j).
\end{equation}
Finally, substitute (\ref{1.19}),  the second equality in
(\ref{1.20}), and (\ref{1.21}), into (\ref{1.10''}) to get
(\ref{1.15}).
\end{proof}


\begin{thebibliography}{AGKS}



\bibitem{CSoS}
P.L. Christiansen (ed.), M.P. S\"orensen (ed.), and A.C. Scott (ed.)
{\it Nonlinear science at the dawn of the 21st century},
Lecture Notes in Physics, V.542, Springer, Berlin,
2000.

\bibitem{CC}
D.V. Chudnovsky and G.V. Chudnovsky, {\it B\"acklund
transformation as a method of decomposition and reproduction of
two-dimensional nonlinear systems}, Phys. Lett. A  {\textbf
{87}}:7 (1982), 325-329.

\bibitem{C1}
   J. Cieslinski,
{\it An algebraic method to construct the Darboux matrix},
 J. Math. Phys. {\bf 36}:10 (1995), 5670-5706.

\bibitem{TF}
L.D. Faddeev and L.A. Takhtajan, {\it Hamiltonian methods in the
theory of solitons}, Springer Verlag, NY, 1986.



\bibitem{GeH}
F. Gesztesy and H. Holden,  {\it Soliton equations and their
algebro-geometric solutions}, Cambridge Studies in Advanced
Mathematics {\bf 79}, Cambridge University Press, Cambridge, 2003.



\bibitem{GKS6}
I. Gohberg, M.A. Kaashoek and A.L. Sakhnovich, {\it Scattering
problems for a canonical system with a pseudo-exponential
potential}, Asymptotic Analysis, {\bf 29}:1 (2002), 1-38.

\bibitem{Gu}
C.H. Gu, H. Hu, and Z. Zhou, {\it  Darboux transformations in
integrable systems}, Springer Verlag, 2005.



\bibitem{KaSa}
M.A. Kaashoek and A.L. Sakhnovich: {\it Discrete skew self-adjoint
canonical system and the isotropic Heisenberg magnet model},  J.
Funct. Anal. {\bf 228} (2005), 207-233.

\bibitem{KePS}
Kevrekidis, P.G.; Pelinovsky, D.E.; Stefanov, A. Nonlinearity
management in higher dimensions. (English) J. Phys. A, Math. Gen.
39, No.3, 479-488 (2006).

\bibitem{Ku}
 Kudryashov, Nikolai A.
Simplest equation method to look for exact solutions of nonlinear
differential equations. (English) Chaos Solitons Fractals 24,
No.5, 1217-1231 (2005).



\bibitem{Mar}
V.A. Marchenko, {\it Nonlinear equations and operator algebras},
Reidel Publishing Co., Dordrecht, 1988.



\bibitem{MS}
V.B. Matveev and M.A. Salle, {\it Darboux transformations and
solitons}, Springer Verlag, Berlin, 1991.

\bibitem{MiG}
 Mieck, B.; Graham, R.
Bose-Einstein condensate of kicked rotators. (English) J. Phys. A,
Math. Gen. 37, No.44, L581-L588 (2004).

\bibitem{Mi}
R. Miura  (ed.), {\it B\"acklund  Transformations}, Lecture Notes
in  Math.  Vol. 515, Springer-Verlag,  Berlin, 1976.





\bibitem{RoS}
C. Rogers and W.K. Schief, {\it B\"acklund and Darboux
transformations. Geometry and modern applications in soliton
theory},  Cambridge Texts in Applied Mathematics, Cambridge
University Press, Cambridge, 2002.

 \bibitem{SaA1}
 A.L. Sakhnovich, {\it Exact solutions of nonlinear equations
and the method of operator identities}, {Lin. Alg. Appl.} { \bf
182} (1993), 109-126.





\bibitem{SaA4}
 A.L. Sakhnovich, {\it Iterated  B\"acklund-Darboux transformation
and transfer matrix-function (nonisospectral case)}, { Chaos,
Solitons and Fractals} { \bf 7}  (1996), 1251-1259.



\bibitem{SaA6} A.L. Sakhnovich, {\it Generalized
B\"acklund-Darboux transformation: spectral properties and
nonlinear equations}, JMAA {\bf 262} (2001), 274-306.



\bibitem{SaA10}
A.L. Sakhnovich, {\it  Matrix Kadomtsev-Petviashvili equation:
matrix identities and explicit non-singular solutions}, J. Phys. A
{\bf 36} (2003), 5023-5033.

\bibitem{SaA11}
A.L. Sakhnovich, {\it Second harmonic generation: Goursat problem
on the semi-strip, Weyl functions and explicit solutions}, Inverse
Problems {\bf 21}:2 (2005), 703-716.

\bibitem{SZ} A.L. Sakhnovich  and J.P. Zubelli,
{\it Bundle bispectrality for matrix differential equations}, IEOT
{\bf 41} (2001), 472-496.

\bibitem{SaL3}
L.A. Sakhnovich,  {\it Spectral theory of canonical differential
systems, method of operator identities}, Operator Theory: Adv.
Appl. {\bf 107}, Birkh{\"a}user Verlag,  Basel-Boston, 1999.

\bibitem{SuSu}
Sulem, Catherine; Sulem, Pierre-Louis The nonlinear Schr\"odinger
equation. Self-focusing and wave collapse. (English) Applied
Mathematical Sciences. 139. New York, NY: Springer.  (1999).



\bibitem{ZM}
V.E.Zakharov and A.V.Mikhailov, {\it On the integrability of
classical spinor models in two-dimensional space-time}, Comm.
Math. Phys. {\bf 74} (1980), 21-40.
\end{thebibliography}
\end{document}